# Conditioning Medicine

www.conditionmed.org

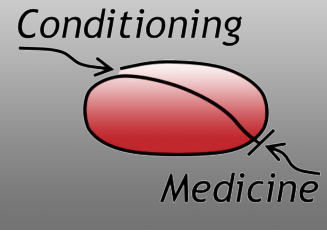

**REVIEW ARTICLE | OPEN ACCESS**

# A new pharmacological preconditioning-based target: from drosophila to kidney transplantation

Michel Tauc[1*], Nicolas Melis[1*], Miled Bourourou[2], Sébastien Giraud[3], Thierry Hauet[3], and Nicolas Blondeau[2]

One of the biggest challenges in medicine is to dampen the pathophysiological stress induced by an episode of ischemia. Such stress, due to various pathological or clinical situations, follows a restriction in blood and oxygen supply to tissue, causing a shortage of oxygen and nutrients that are required for cellular metabolism. Ischemia can cause irreversible damage to target tissue leading to a poor physiological recovery outcome for the patient. Contrariwise, preconditioning by brief periods of ischemia has been shown in multiple organs to confer tolerance against subsequent normally lethal ischemia. By definition, preconditioning of organs must be applied preemptively. This limits the applicability of preconditioning in clinical situations, which arise unpredictably, such as myocardial infarction and stroke. There are, however, clinical situations that arise as a result of ischemia-reperfusion injury, which can be anticipated, and are therefore adequate candidates for preconditioning. Organ and more particularly kidney transplantation, the optimal treatment for suitable patients with end stage renal disease (ESRD), is a predictable surgery that permits the use of preconditioning protocols to prepare the organ for subsequent ischemic/reperfusion stress. It therefore seems crucial to develop appropriate preconditioning protocols against ischemia that will occur under transplantation conditions, which up to now mainly referred to mechanical ischemic preconditioning that triggers innate responses. It is not known if preconditioning has to be applied to the donor, the recipient, or both. No drug/target pair has been envisioned and validated in the clinic. Options for identifying new target/drug pairs involve the use of model animals, such as drosophila, in which some physiological pathways, such as the management of oxygen, are highly conserved across evolution. Oxygen is the universal element of life existence on earth. In this review we focus on a very specific pathway of pharmacological preconditioning identified in drosophila that was successfully transferred to mammalian models that has potential application in human health. Very few mechanisms identified in these model animals have been translated to an upper evolutionary level. This review highlights the commonality between oxygen regulation between diverse animals.

<>
[1]Université Côte d'Azur, CNRS UMR-7370, LP2M, Nice, F-06107, France. [2]Université Côte d'Azur, CNRS UMR-7275, IPMC, Sophia Antipolis, F-06560, France. [3]Université de Poitiers, Faculté de Médecine et de Pharmacie, Inserm U1082, Poitiers, F-86000, France.

*These authors participated equally to this work.
Correspondence should be addressed to Michel Tauc (michel.tauc@unice.fr).

**Introduction**

Organ transplantation is characterized by an unavoidable episode of ischemia/reperfusion (IR), which is one of the main causes for impairment in graft functional recovery and poor patient outcome. Kidney transplantation has become an established worldwide practice replacing the endless dialysis strategy that negatively affects a patient's well-being and finances. For many years clinicians have been looking for any way to reduce the ischemia induced renal stress that occurs after reperfusion, which decreases the optimal functional recovery of the graft (Khalifeh et al., 2015). Unfortunately, very few protocols addressing this particular ischemic situation have been transferred to the human clinic. This lack of clinical translation is probably due to the multifactorial causes of the ischemia/reperfusion-induced injury.

While ischemic injuries of the brain and heart are unpredictable in their onset, organ transplantation is a scheduled surgical procedure. The predictability of organ transplantation therefore facilitates the use of protective preconditioning strategies either on the donor, the recipient, or on the isolated organ. It should however be borne in mind that while preconditioning is logistically straightforward in the case of elective living donor kidney transplantation, in deceased donor transplantation, preconditioning is much more of a challenge as the timing is unpredictable and the donor and recipient are often in different hospitals. In this situation the most effective protocol often tested at the preclinical level is ischemic preconditioning (IPC), as applied on the kidney (Toosy et al., 1999). Basically the principle of this method is to apply brief repeated occlusions of the arteries irrigating the organ in order to induce a protective response against a subsequent prolonged potentially lethal occlusion. Derived from IPC, remote ischemic preconditioning (RIPC) confers a global protective effect and renders remote organs resistant to IR injury through various transduction signals, mediators, and effectors that have been recently reviewed (Hausenloy et al., 2016; Stokfisz et al., 2017). Despite the promising results described in animal models (Torras et al., 2002; Soendergaard et al., 2012; Yoon et al., 2015) it is difficult to draw clear conclusions regarding the potential benefit of ischemic conditioning in clinical trials (MacAllister et al., 2015; Nadarajah et al., 2017; Veighey, 2018). Various IPC and RIPC protocols have been used in these clinical trials either on the donor or the recipient. As the mechanism underlining the preconditioning effect has not yet been fully elucidated it is more difficult to plan the most effective and appropriate intervention in the clinical setting. A review of the existing clinical trials literature in this area highlights the potential of RIPC in kidney transplantation (Veighey and MacAllister, 2017). The recent REPAIR clinical trial (MacAllister et al., 2015), which is currently the largest clinical study, has provided evidence that RIPC improves the glomerular filtration rate after kidney transplantation. These encouraging results have motivated clinicians to look for pharmacological conditioning agents (preconditioners), which reproduce the benefit conferred by RIPC. Model organisms, such as drosophila, could help uncover new paradigms and new pathways involved in ischemic protection, possibly leading to innovative therapeutic strategies in the future. The use of animal models to examine conditioning strategies against ischemic injury is reasonable given that almost all organisms on earth live by virtue of the oxygen. Thus it could be reasonably hypothesized that evolution could have selected some common or similar molecular mechanisms concerning the management of oxygen restriction between species.

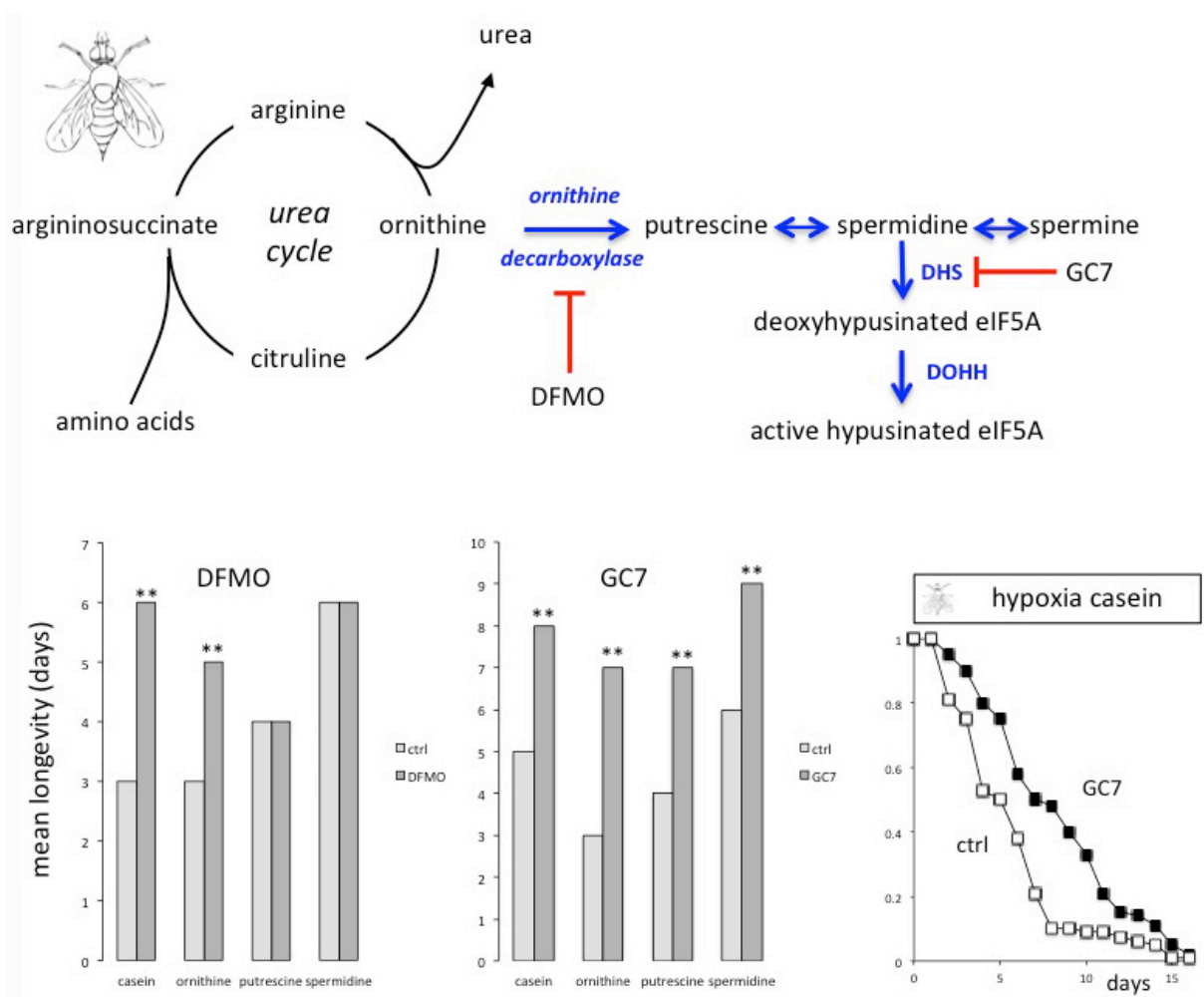

Figure 1. Schema of the urea cycle involved in the catabolism of amino acids leading to polyamines synthesis through sequential enzymatic catalysis. Survivorship of flies maintained under hypoxia depends on the amino acid content of the diet. Inhibition of ornithine decarboxylase with DFMO suppresses the toxic effects of a casein diet or of ornithine added to a glucose protein free diet on hypoxic flies' longevity. The polyamines putrescine and spermidine remained deleterious in the presence of DFMO. By contrast inhibition of DHS with GC7 strongly prevented the deleterious actions of casein, amino acids, and polyamines on flies' survival under hypoxic conditions. The survivorship of hypoxic flies maintained on a casein diet was enhanced by 72% in the presence of GC7. A log rank test was used to compare survivorship curves. **: $p < 0.0001$. DHS: deoxyhypusine synthase, DOHH: deoxyhypusine synthase, DFMO: α-difluoromethylornithine, GC7: N1-guanyl-1,7-diaminoheptane. From Vigne and Frelin (Vigne and Frelin, 2008).

**Emergence of a new concept in hypoxic tolerance**
The team of Dr. C. Frelin was a pioneer in this matter and deciphered in drosophila a unique pathway involved in hypoxic tolerance. Starting from their own data demonstrating that restriction of dietary proteins dramatically increased the longevity of hypoxic flies (Vigne and Frelin, 2006; 2007; Vigne et al., 2009), they succeeded in identifying the mechanism involved in the diet dependent hypoxic tolerance. First, by exposing the flies to a 10% sucrose diet supplemented with an incremental concentration load of proteins by the addition of yeast or casein, they observed a dose dependent inhibitory action of yeast and casein on the longevity of flies maintained in hypoxic conditions (Vigne and Frelin, 2008). Then trying to understand which amino acid was responsible for the negative effect on survival to hypoxia, they found that any one of the natural amino acids could reproduce the action of yeast or casein at the millimolar level. This last observation suggested that amino acids act through a more general metabolic pathway. Since the common catabolic amino-acid pathway leading to nitrogen wasting is the urea cycle (Figure 1), Vigne and Frelin (2008) further examined the possible involvement of this pathway in hypoxic tolerance. They supplemented the 10% sucrose diet with L-citrulline or L-ornithine, 2 amino acids intermediates of the urea cycle, which are not incorporated into proteins. Subjecting flies fed this diet to hypoxic challenges revealed that both L-citrulline and L-ornithine reproduced the effects of individual amino acids confirming the involvement of the urea cycle in the susceptibility to hypoxia. Since L-ornithine is known to feed the polyamine synthesis pathway, they then tested the effect of the polyamines putrescine, spermine, and spermidine and showed that when added to the sucrose diet, these polyamines also decreased the mean longevities of hypoxic flies.

As highlighted in Figure 1, the first step in the synthesis of polyamines is the decarboxylation of L-ornithine to give putrescine via the action of ornithine decarboxylase (ODC) (Kahana, 2018). α-Difluoromethylornithine (DFMO), a specific inhibitor of ODC (LoGiudice et al., 2018), increased the median and maximum longevities of hypoxic flies fed the deleterious diet enriched in casein, L-asparagin, or L-ornithine. As DFMO reversed the life-shortening effect of amino acids, it confirmed the involvement of polyamines. This involvement was substantiated by the fact that DFMO could not preserve the longevity of hypoxic flies fed a sucrose diet supplemented with putrescine or spermidine (Vigne and Frelin, 2008). Polyamines exert numerous and essential cell functions (Pegg, 2016), including the activation of the eukaryotic translation initiation factor eIF5A through the addition of hypusine, which is transferred from spermidine to a specific lysine moiety via the catalytic activity of deoxyhypusine synthase (DHS) and the subsequent action of deoxyhypusine hydroxylase (DOHH) (Figure 1) (Park, 2006). By using the specific DHS competitive inhibitor N1-guanyl-1,7-diaminoheptane (GC7) Vigne and Frelin (2008) showed that inhibition of DHS dramatically enhanced the longevity of hypoxic flies fed a diet containing amino acids or polyamines. Overall, the conclusion drawn from their research was that inhibition of eIF5A hypusination is crucial for promoting hypoxic tolerance (Figure 1).

**A successful translation of the concept rose from a model organism to an integrative mammalian model of ischemia**
eIF5A is the only known protein to contain the unusual amino acid hypusine [N (ε)- (4-amino-2-hydroxybutyl)-lysine],

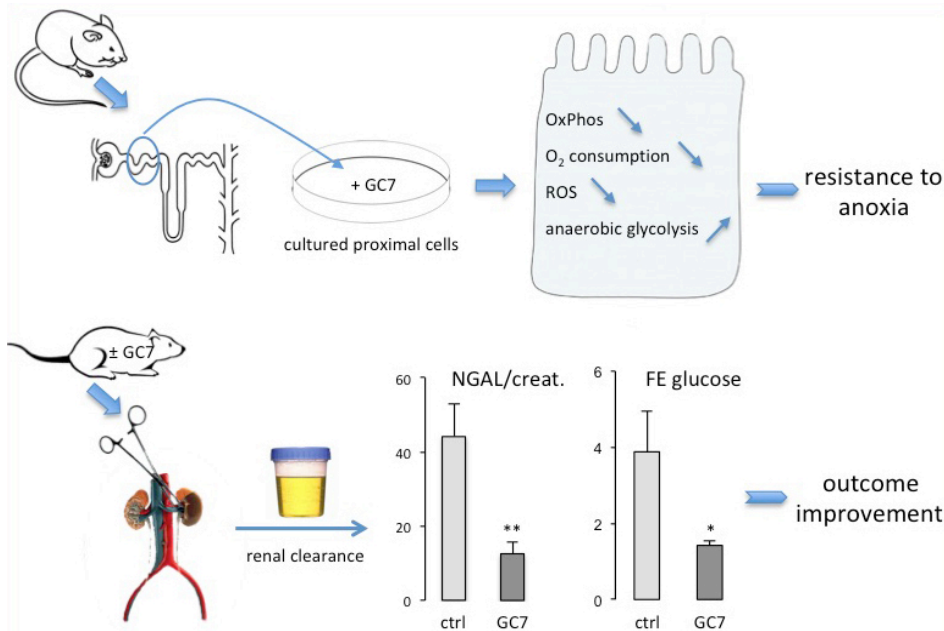

Figure 2: Schema of the translation of the concept that arose from a model organism to an integrative mammalian model of ischemia. *In vitro* analysis of the effects of GC7 was performed on mouse cultured proximal cells. Preconditioning with GC7 provided tolerance to anoxia accompanied by a fall in oxidative phosphorylation (OxPhos), a fall in oxygen consumption, a fall in reactive oxygen species (ROS) generation, and by a metabolic switch toward anaerobic glycolysis to maintain the energetic status of the cells. *In vivo* analysis was performed on a unilateral model of kidney ischemia in rats. A 40 min blood flux arrest induced an alteration of renal function that was largely prevented at the structural (NGAL) and functional level (FE glucose) by preconditioning the animals with GC7. Reproduced from Melis et al. (2017).

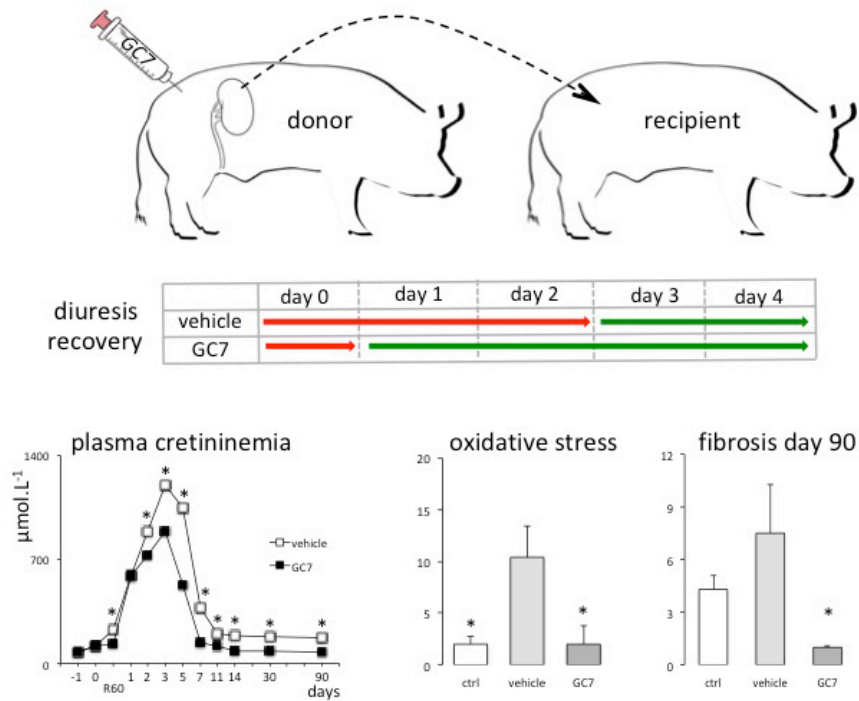

Figure 3: Schema of the preclinical use of GC7 in kidney transplantation. The donor pig was preconditioned with GC7 before kidney removal. Preservation and transplantation followed a classical protocol used in the human clinic. GC7 preconditioned animals recovered a normal diuresis earlier than control animals. The plasma creatinine level in GC7 preconditioned animals remained below control animals up to 3 months post transplantation confirmed the protective effect of GC7 on the graft. GC7 also prevented the increase in the oxidative stress level and the development of fibrosis. R60 is the time 60 min after reperfusion. Reproduced from Melis et al. (2017).

which is synthetized on eIF5A at a specific lysine residue from the polyamine spermidine. eIF5A, DHS, and DOHH are found in all eukaryotes and are highly conserved through evolution, suggesting a crucial function of eIF5A and the deoxyhypusine/hypusine modification (Park, 2006). It therefore seemed interesting to take advantage of the universal need in animal species for oxygen to try to translate and expand the data obtained in drosophila to a more integrated mammalian system. For this matter, the choice of kidney as a reporter organ either *in vitro* or *in vivo* seemed relevant since its functional integrity is highly dependent on the oxygen level (Hansell et al., 2013). For the first time using cultured cells from mouse proximal tubule (Duranton et al., 2012) subjected to strong hypoxia, we showed that GC7, as well as siRNA targeting DHS or DOHH, was able to largely prevent the anoxia induced cell death. This tolerance to ischemia was shown to follow a fall in oxygen consumption originating from a reversible oxidative phosphorylation (OxPhos) shutdown with the energetic status being maintained by a metabolic shift toward anaerobic glycolysis (Melis et al., 2017). The main consequence of this mitochondria "silencing" was the prevention of reactive oxygen species (ROS) generation. Based on these results, the authors performed a functional analysis of renal function in rats whose kidneys were subjected to the classical unilateral ischemia/reperfusion challenge (Kato et al., 2014). GC7 preconditioning of these rats induced a drastic reduction in the hypusinated form of eIF5A as shown by western blot analysis of kidneys. Twenty-four hours after the ischemia/reperfusion challenge, the main renal parameters were checked by individual clearance analysis and compared in each animal to those obtained from the contralateral non-ischemic kidney (Melis et al., 2017). Concentrations of urinary neutrophil gelatinase-associated lipocalin (NGAL), an early marker of renal injury, significantly increased in the ischemic kidney together with an increase in sodium, glucose, and phosphate fractional excretion (FEs). GC7 preconditioning prevented the increase in urine levels of NGAL. Furthermore, GC7 largely protected proximal tubular function, since the FEs of glucose, $PO_4^{3-}$, and $Na^+$ remained similar to those observed in the control non-ischemic kidney (Figure 2). These observations identified an innovative HIF-independent pathway involved in hypoxic and ischemic tolerance that originally emerged from the drosophila model.

### The perspective of a clinical translation in kidney transplantation

As an experimental model, transplantation is obviously the best for predictable ischemic injury and evaluation of new protective concepts. From a clinical point of view, due to the success and the progress in transplantation science, the demand for organs continuously increases leading to longer wait time on transplantation waiting lists. This issue could be partially overcome through the extension of donor criteria and the acceptance of marginal donors (Rosengard et al., 2002). Organs, specifically kidneys from this specific population are particularly sensitive to ischemia reperfusion injury and up to now have been eliminated from the pool of available organs. Improving or innovating preconditioning protocols will enhance this pool by favorably broadening the inclusion criteria. To explore this kind of concept we tested the use of GC7 in a relevant preclinical model of kidney transplantation in pigs (Giraud et al., 2011; Melis et al., 2017). In this model, the ischemia reperfusion alone induced early and late graft dysfunction, whereas preconditioning of the donor with GC7 improved graft functional recovery and late graft function as assessed by measurements of creatininemia, sodium FE, osmolarity plasma/urine, alanine aminopeptidase excretion, and plasma aspartate aminotransferase level. A major finding was the recovery of diuresis that lasted 3 days in vehicle treated animals whereas it was effective one day post-transplantation in GC7-preconditioned pigs. Additionally the use of GC7 lowers the oxidative stress measured on biopsies post-surgery. At longer time points GC7 efficiently protected against the development of chronic interstitial fibrosis 3 months post-transplantation. Data also showed that GC7 exposure had a

long–term protective effect on the graft against the inhibition of fibrinolysis in the vessels known to promote fibrosis. Furthermore the expression of the fibrosis–related factor (Brigstock, 2010), connective tissue growth factor (CTGF), was suppressed in the GC7-preconditioned group. It appears clearly that GC7 preconditioning improves graft function outcome in kidney transplantation (Figure 3).

**Conclusion**

It has been demonstrated that the intensity of ischemic-reperfusion injury very strongly correlates with delayed graft function, chronic graft dysfunction, and late graft loss. This situation places ischemic injury at the forefront of issues to be addressed in transplantation. Preconditioning is evidently an attractive and promising way to prepare the candidate organ for the harmful stress of oxygen deprivation (Veighey and MacAllister, 2017). A protocol of ischemic preconditioning consists of the application of brief nonlethal periods of ischemia that activate an innate response that confers protection against future potentially lethal periods of ischemia (Veighey and MacAllister, 2017; Veighey, 2018). Deriving from the original concept of ischemic preconditioning, various conditioning paradigms including pharmacological preconditioning may be promising as innovative therapies for prevention of ischemic-related injury like transplantation. The data reported in the literature shows that GC7 is an efficient cell and organ preconditioner against ischemia. GC7-preconditioning involves the inhibition of the activation step of eIF5A that appears to be a promising target for organ transplantation. Observed first in drosophila, targeting this innovative target is important for conferring ischemic tolerance in mammals and by extension, of potential use in clinical situations in which a fall in oxygen delivery is involved. This could serve as a basis from which to launch further investigation into therapies for organ transplantation or any other predictable ischemia related injuries. One future avenue of examination in to preconditioning strategies (Tauskela and Blondeau, 2018) would be selecting combinations of therapies, including different preconditioning protocols that likely act through different pathways to obtain synergistic action with a better outcome of graft function.

**Acknowledgments**

The authors thank all their past and present team members and collaborators who have contributed to the data discussed in the review.

**Financial support and sponsorship**

This work was supported in part by the Centre National de la Recherche Scientifique (CNRS), grant ANR-08-GENO-022 from the Agence Nationale de la Recherche, grant DPM 20121125559 from the Fondation pour la recherche médicale (FRM), and a grant from the Société d'Accélération de Transfert de Technologie SATT Sud-Est.

**Conflict of interest statement**

The authors declare that they have no conflicts of interest.

**References**

Brigstock DR (2010) Connective tissue growth factor (CCN2, CTGF) and organ fibrosis: lessons from transgenic animals. J Cell Commun Signal 4:1–4.

Duranton C, Rubera I, Cougnon M, Melis N, Chargui A, Mograbi B, Tauc M (2012) CFTR is involved in the fine tuning of intracellular redox status: physiological implications in cystic fibrosis. Am J Pathol 181:1367–1377.

Giraud S, Favreau F, Chatauret N, Thuillier R, Maiga S, Hauet T (2011) Contribution of large pig for renal ischemia-reperfusion and transplantation studies: the preclinical model. J Biomed Biotechnol 2011:532127.

Hansell P, Welch WJ, Blantz RC, Palm F (2013) Determinants of kidney oxygen consumption and their relationship to tissue oxygen tension in diabetes and hypertension. Clin Exp Pharmacol Physiol 40:123–137.

Hausenloy DJ et al. (2016) Ischaemic conditioning and targeting reperfusion injury: a 30 year voyage of discovery. Basic Res Cardiol 111:70.

Kahana C (2018) The antizyme family for regulating polyamines. J Biol Chem 293:18730–18735.

Kato J, Nakayama M, Zhu W-J, Yokoo T, Ito S (2014) Ischemia/reperfusion of unilateral kidney exaggerates aging-induced damage to the heart and contralateral kidney. Nephron Exp Nephrol 126:183–190.

Khalifeh T, Baulier E, Le Pape S, Kerforne T, Coudroy R, Maiga S, Hauet T, Pinsard M, Favreau F (2015) Strategies to optimize kidney recovery and preservation in transplantation: specific aspects in pediatric transplantation. Pediatr Nephrol 30:1243–1254.

LoGiudice N, Le L, Abuan I, Leizorek Y, Roberts SC (2018) Alpha-Difluoromethylornithine, an Irreversible Inhibitor of Polyamine Biosynthesis, as a Therapeutic Strategy against Hyperproliferative and Infectious Diseases. Med Sci (Basel) 6; doi:10.3390/medsci6010012.

MacAllister R, Clayton T, Knight R, Robertson S, Nicholas J, Motwani M, Veighey K (2015) REmote preconditioning for Protection Against Ischaemia–Reperfusion in renal transplantation (REPAIR): a multicentre, multinational, double-blind, factorial designed randomised controlled trial. Southampton (UK): NIHR Journals Library. Efficacy Mech Eval 2; doi: 10.3310/eme02030..

Melis N, Rubera I, Cougnon M, Giraud S, Mograbi B, Belaid A, Pisani DF, Huber SM, Lacas-Gervais S, Fragaki K, Blondeau N, Vigne P, Frelin C, Hauet T, Duranton C, Tauc M (2017) Targeting eIF5A Hypusination Prevents Anoxic Cell Death through Mitochondrial Silencing and Improves Kidney Transplant Outcome. J Am Soc Nephrol 28:811–822.

Nadarajah L, Yaqoob MM, McCafferty K (2017) Ischemic conditioning in solid organ transplantation: is it worth giving your right arm for? Curr Opin Nephrol Hypertens 26:467–476.

Park MH (2006) The post-translational synthesis of a polyamine-derived amino acid, hypusine, in the eukaryotic translation initiation factor 5A (eIF5A). J Biochem 139:161–169.

Pegg AE (2016) Functions of Polyamines in Mammals. J Biol Chem 291:14904–14912.

Rosengard BR, Feng S, Alfrey EJ, Zaroff JG, Emond JC, Henry ML, Garrity ER, Roberts JP, Wynn JJ, Metzger RA, Freeman RB, Port FK, Merion RM, Love RB, Busuttil RW, Delmonico FL (2002) Report of the Crystal City Meeting to Maximize the Use of Organs Recovered from the Cadaver Donor. American Journal of Transplantation 2:701–711.

Soendergaard P, Krogstrup NV, Secher NG, Ravlo K, Keller AK, Toennesen E, Bibby BM, Moldrup U, Ostraat EO, Pedersen M, Jorgensen TM, Leuvenink H, Norregaard R, Birn H, Marcussen N, Jespersen B (2012) Improved GFR and renal plasma perfusion following remote ischaemic conditioning in a porcine kidney transplantation model. Transpl Int 25:1002–1012.

Stokfisz K, Ledakowicz-Polak A, Zagorski M, Zielinska M (2017) Ischaemic preconditioning – Current knowledge and potential future applications after 30 years of experience. Adv in Med Sci 62:307–316.

Tauskela JS, Blondeau N. (2018) Requirement for preclinical

prioritization of neuroprotective strategies in stroke: Incorporation of preconditioning. Conditioning Medicine 1(3): 124-134.

Toosy N, McMorris EL, Grace PA, Mathie RT (1999) Ischaemic preconditioning protects the rat kidney from reperfusion injury. BJU Int 84:489–494.

Torras J, Herrero-Fresneda I, Lloberas N, Riera M, Ma Cruzado J, Ma Grinyó J (2002) Promising effects of ischemic preconditioning in renal transplantation. Kidney Int 61:2218–2227.

Veighey K (2018) Ischemic conditioning in organ transplantation. conditioning Med 1:212–219.

Veighey K, MacAllister R (2017) Ischemic Conditioning in Kidney Transplantation. J Cardiovasc Pharmacol Ther 22:330–336.

Vigne P, Frelin C (2006) A Low Protein Diet Increases the Hypoxic Tolerance in Drosophila. PLoS One 1: e56 doi:10.1371/journal.pone.0000056.

Vigne P, Frelin C (2007) Diet dependent longevity and hypoxic tolerance of adult Drosophila melanogaster. Mech Ageing Dev 128:401–406.

Vigne P, Frelin C (2008) The role of polyamines in protein-dependent hypoxic tolerance of Drosophila. BMC Physiol 8:22.

Vigne P, Tauc M, Frelin C (2009) Strong dietary restrictions protect Drosophila against anoxia/reoxygenation injuries. PLoS One 4:e5422; doi:10.1371/journal.pone.0005422.

Yoon YE, Lee KS, Choi KH, Kim KH, Yang SC, Han WK (2015) Preconditioning strategies for kidney ischemia reperfusion injury: implications of the "time-window" in remote ischemic preconditioning. PLoS ONE doi:10:e0124130.